\documentclass[12pt,preprint]{aastex}

\usepackage{natbib}

\slugcomment{}

\begin{document}

\title{A New CEMP-s RR Lyrae Star }

\author{T.D. Kinman\altaffilmark{1}          }
\affil{National Optical Astronomy Observatory, 950 N. Cherry Avenue,
Tucson, AZ 85719, USA}

\author{Wako Aoki\altaffilmark{2}}                             
\affil{National Astronomical Observatory of Japan, Mitaka, Tokyo
181-8588, Japan}

\author{Timothy C. Beers\altaffilmark{1}}
\affil{National Optical Astronomy Observatory, 950 N. Cherry Avenue,
Tucson, AZ 85719, USA}

\author{ Warren  R. Brown}                             
\affil{Smithsonian Astrophysical Observatory, 60 Garden St.,    
 Cambridge, MA 02138, USA}

\altaffiltext{1}{ 
NOAO is operated by AURA, Inc.\ under cooperative 
agreement with the National Science Foundation.}
\altaffiltext{2}{
Department of Astronomical Science, School of Physical Sciences, 
The Graduate University of Advanced Studies (SOKENDAI), 2-21-1 Osawa,
Mitaka, Tokyo 181-8588, Japan}

\begin{abstract}

We show that SDSS J170733.93+585059.7 (hereafter SDSS J1707+58),
previously identified by Aoki and collaborators as a carbon-enhanced
metal-poor star (with s-process-element enhancements; CEMP-s), on the
assumption that it is a main-sequence turn-off star, is the RR Lyrae
star VIII-14 identified by the Lick Astrograph Survey. Revised
abundances for SDSS J1707+58 are [Fe/H] = --2.92, [C/Fe] = +2.79, and
[Ba/Fe] = +2.83. It is thus one of the most metal-poor RR Lyrae stars
known, and has more extreme [C/Fe] and [Ba/Fe] than the only other RR
Lyrae star known to have a CEMP-s spectrum (TY Gru). Both stars are
Oosterhoff II stars with prograde kinematics, in contrast to stars with
[C/Fe]$<+0.7$, such as KP Cyg and UY CrB, which are disk stars. Twelve
other RR Lyrae stars with [C/Fe] $\geq +0.7$ are presented as CEMP
candidates for further study. 
 
\end{abstract}


\section{Introduction}

Stars with strong carbon features, such as molecular CH and C$_2$, in
their spectra are referred to as carbon stars; Wallerstein \& Knapp
(1998) provide a detailed review of carbon-star properties. High-
velocity CH (strong G-band) stars were first noticed by Keenan (1942).
Bond (1974) observed a metal-weak subgroup of these stars whose spectra
exhibited over-abundant CH and s-process elements. He called them
``Population II carbon stars", and showed that they had a wide      
range in luminosities. CH stars can be found among the red giants
in the globular cluster $\omega$ Cen, although they
are rare in other globulars. Stars with these characteristics are now
known to be a very heterogeneous group.

Beers \& Christlieb (2005) defined metal-poor stars as those with [Fe/H]
$<$ --1.0, and carbon-enhanced metal-poor stars with enhanced s-process
elements (referred to by these authors as CEMP-s stars) as those with
[C/Fe] $>$ +1.0 and [Ba/Fe] $>$ +1.0. More recently, Suda et al. (2011)
have considered CEMP-s stars to be those that have [Fe/H] $\leq$ --2.5,
[C/Fe] $\geq$ +0.7, and [Ba/Fe] $\geq$ +0.5 (explicitly including very
low metallicity in their definition), although large numbers of more
metal-rich CEMP-s stars satisfying the original Beers \& Christlieb
criteria certainly exist, and indeed represent the majority of such
objects.

Wallerstein et al. (2009), in an ongoing survey of carbon abundances of
RR Lyrae stars, have shown that the two metal-rich RR Lyrae stars, KP
Cyg and UY CrB, have [C/Fe] of +0.52 and +0.65, respectively. KP Cyg and
UY CrB have long periods for such metal-rich stars, and Andrievsky et
al. (2010) suggest that they are instead short-period Cepheids.
Additionally, Wallerstein \& Huang (2010) have given carbon abundances
for 24 RR Lyrae stars -- all had [C/Fe] $\leq$ +0.15; they concluded
that higher [C/Fe] were found in more metal-rich stars.

The only RR Lyrae star now known to be a CEMP-s star is TY Gru, a Bailey
type ab star with a period of 0.57 days, found by Preston et al. (2006)
among the metal-poor candidates identified during the course of the HK
Survey of Beers and collaborators (CS~22881-071; Beers et al. 1992).
Preston et al. showed that TY Gru has [Fe/H] = --2.0, [C/Fe] = +0.9,
[Ba/Fe] = +1.2, as well as enhancements among other s-process-element
abundances. Stars with such abundances have been shown to be binaries in
which mass transfer has occurred from an AGB companion at an earlier
evolutionary stage. TY Gru, however, has a variable light and velocity
curve (Blazhko effect), and its binary nature has yet to be shown, despite
strenuous efforts by Preston (2009). 

Measurements of carbon abundances among metal-poor stars (e.g., Beers \&
Christlieb 2005, and references therein; Aoki et al. 2012) have
established the increasing frequency of stars with large [C/Fe] with
decreasing [Fe/H]. Carollo et al. (2012) have determined the [C/Fe]
abundance ratio for over 30,000 calibration stars from the Sloan Digital
Sky Survey (SDSS; York et al. 2000). They confirm the increasing
frequency of stars with [C/Fe] $>$ +0.7 (CEMP stars) with decreasing
[Fe/H], and also with increasing distance ($\mid$Z$\mid$) from the
Galactic plane. The frequency of CEMP stars kinematically associated
with the outer halo was also shown by these authors to be roughly twice
that of such stars kinematically associated with the inner halo. For
further background, see Carollo et al. (2010). One of the CEMP stars in
their sample is SDSS J1707+58. Aoki et al. (2008) identified it as a
CEMP star among a number of presumed main-sequence turn-off stars from
the SDSS/SEGUE sample, interpreting its observed rapid variations in
radial velocity as due to motion within a tight binary system.

In this paper, we show that SDSS J1797+58 is the RR Lyrae star VIII-14,
identified in the spectroscopic study of RR Lyrae stars of Suntzeff et
al. (1991), who assigned an [Fe/H] = --2.5, and noted that it had a
strong CH G-band. Aoki et al. (2008) estimated an [Fe/H] = --2.52, along
with [C/Fe] = +2.1 and [Ba/Fe] = +3.40; we revise these values on the
basis of its new identification as an RR Lyrae. The presence of carbon
enhancements in RR Lyrae stars is of particular interest, since they are
excellent population tracers in the Galaxy. In this paper we compare
SDSS J1707+58 with the other carbon-enhanced variables TY Gru, KP Cyg,
and UY CrB. We also discuss how additional samples of such stars might
be found. 

\section {Photometry, Radial Velocities, and Ephemeris of SDSS J1707+58}

The RR Lyrae star VIII-14 (Suntzeff et al. 1991) was originally
discovered in a survey with the Lick Astrograph (Kinman et al. 1984);
the star is in field VIII \footnote{ A paper describing the stars in
this field is currently being prepared for publication.}, and has
coordinates 17$^{h}$ 07$^{m}$ 33.$^{s}$93 +58$^{\circ}$ 50$^{'}$
59.$^{"}$9 (J2000) and Galactic coordinates $l =$ 87.$^{\circ}$69, $b =
$+36.$^{\circ}$29. Its proper motion, given in SDSS DR8
(Adelman-McCarthy et al. 2011), is $\mu_{\alpha}$ = --4$\pm$3 mas
y$^{-1}$ and $\mu_{\delta}$ = +3$\pm$3 mas y$^{-1}$. We derive
photographic B magnitudes from 42 exposures on 103-aO emulsion (May 1964
to August 1965; blue open circles in Fig 1c), and from 32 exposures on
II-aO emulsion (June 1976 to April 1985; red filled circles in Fig 1c),
using the Lick Carnegie Astrograph \footnote {For further details see
Kinman (1965).}. The exposures on the latter plates are longer, and the
magnitudes derived from them should be more accurate. We derive the B
magnitudes of the comparison stars from their SDSS $ugr$ magnitudes,
using the transformations of Lupton (2005). Photoelectric $BV$
magnitudes of VIII-14 were obtained with the KPNO 1.3-m telescope
\footnote{ For more details see Sec. 3.1 of Kinman et al. (1994).}
between 1979 March 25 UT and 1986 June 10 UT (black filled circles in
Fig 1b and 1c). The radial velocities are taken from Table 3 of Aoki et
al. (2008), and phased according to the method of Liu (1991), assuming a
$V$ amplitude of 0.61 mag. This yields a $\gamma$-velocity of +18.5 km
s$^{-1}$, and a JD hel (at maximum) at 2,454,141.503. From this we
derive the following ephemeris for data between 1964 and 2007:
\begin{eqnarray}  
   JD(max)hel = 2438526.276 + 0.6788049 \times E    \nonumber 
\end{eqnarray}  
The phases used in Fig. 1 are derived from this ephemeris. The second
set of photographic data (red filled circles) may be slightly shifted in
phase with respect to the first set (blue open circles). The shift is
equivalent to a period increase of $\sim$0.14 d Myr$^{-1}$, with an error
of the same order (see La Borgne et al. 2007). 

\begin{figure}
\includegraphics[width = 18.0 cm]{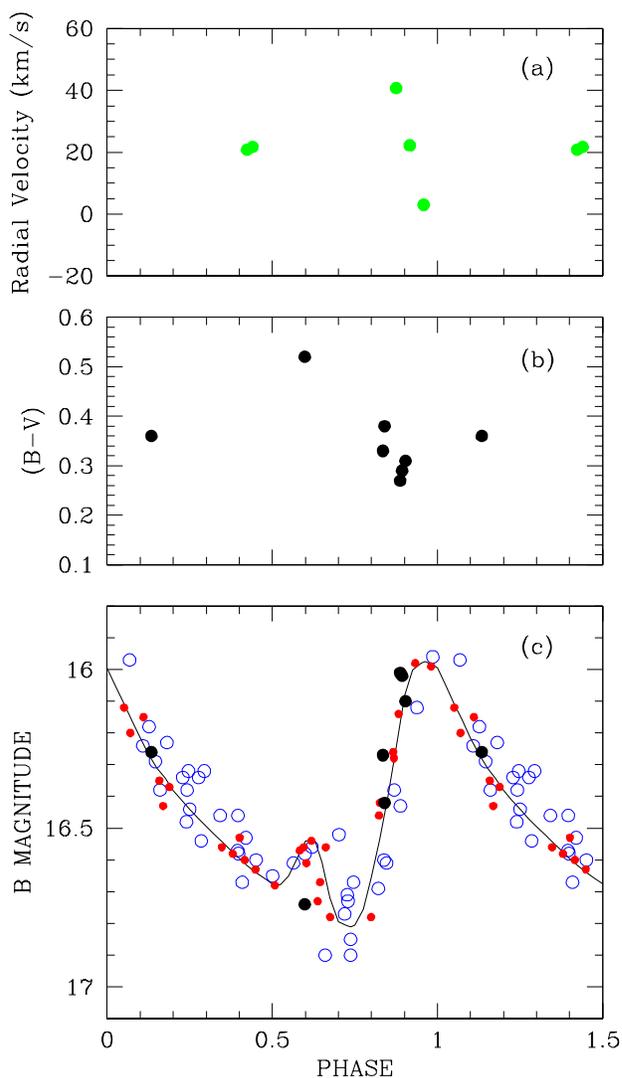}
\caption {(a) Heliocentric radial velocities of SDSS J1707+58, from Table 3
of Aoki et al. (2008), as a function of phase. (b) Photoelectric $(B-V)$
observations of SDSS J1707+58, as a function of phase. (c) $B$
magnitudes of SDSS J1707+58, as a function of phase. The black filled
circles are the photoelectric $B$ magnitudes, the red filled symbols are
photographic photometry obtained between May 1964 and August 1965, and
the blue open circles are photographic photometry obtained between June
1976 and April 1985. }
\label{Fig1}
\end{figure}
 
\section{Revised Chemical Abundances for SDSS J1707+58}

The identification of SDSS J1707+58 as an RR Lyrae star requires a
revision of the abundance analysis of Aoki et al. (2008). We excluded
the spectra obtained at $\phi \sim 0.9$, where the variation with phase
of the spectral lines is greatest. We adopted the two spectra with $\phi
\sim 0.4$ for analysis. Their total exposure time is 30 minutes; this
yields a S/N ratio of 40 per resolution element at 5200~{\AA}. The
effective temperature is estimated from the $(B-V)$ color of 0.43 at
this phase (Fig 1b). Allowing for the reddening ($E(B-V) = 0.028$;
Schlafly \& Finkbeiner 2011), the effective temperature is estimated to
be 6100 -- 6400~K, using the temperature scales of Alonso et al. (1999)
and Casagrande et al. (2010). We adopt 6250~K for the abundance
analysis. This is significantly lower than the effective temperature
adopted by Aoki et al. (2008), who assumed that the object is a turn-off
star.

Our abundance analysis used the ATLAS/NEWODF model
atmospheres (Castelli \& Kurucz 2003). The microturbulent velocity,
$v_{\rm mic}$, is not well-determined from our analysis of the Fe
lines, because the number of the available \ion{Fe}{1} lines is small.
We assume $v_{\rm mic}$ = 3.5~km~s$^{-1}$, with an uncertainty of
1.0~km~s$^{-1}$, taking into account the values assumed in previous
studies of very metal-poor RR Lyrae stars: 4.1~km~s$^{-1}$ (Preston et
al. 2006), 3.0~km~s$^{-1}$ (Hansen et al. 2011). We find no significant
correlation between the equivalent widths of the \ion{Fe}{1} lines and
the Fe abundances derived from the individual lines in our analysis,
which assumes $v_{\rm mic}$ = 3.5~km~s$^{-1}$.

The surface gravity ($g$ [cm s$^{-2}$]) is determined by requiring that
the results from the two ionization species \ion{Fe}{1} and \ion{Fe}{2}
agree. Our estimate, $\log g = 2.8$, although uncertain because of the
limited number of Fe lines, is a reasonable value for very metal-poor RR
Lyrae stars (Hansen et al. 2011).

The new abundances are listed in Table 1. The abundance errors are
estimated in the same way as by Aoki et al. (2008). Abundance errors
caused by uncertainties in the atmospheric parameters are calculated for
$\delta(T_{\rm eff}) = 200$~K, $\delta (\log g) = 0.5$~dex, and
$\delta(v_{\rm mic}) = 1.0~$km~s$^{-1}$. The iron abundance is
significantly lower than that found by Aoki et al., (2008), because we
adopt a lower $T_{\rm eff}$ and higher $v_{\rm mic}$ in the present
analysis. SDSS J1707+58 is one of the most metal (iron)-poor RR Lyrae
stars known to date (see Hansen et al. 2011). The C abundance is
determined from the CH molecular band at 4323~{\AA}. Although the result
is uncertain because of the limited S/N at this wavelength, the large
enhancement of C with respect to Fe is remarkable. The excesses of Na,
Mg, Sr, and Ba reported by Aoki et al. (2008) are also found in this new
analysis. The Ba abundance given by Aoki et al. (2008) is significantly
higher than the solar value ([Ba/H]$= +0.88$); that found in the present
analysis is smaller ([Ba/H]$= -0.09$). The enhancements of C and Ba
relative to Fe in SDSS J1707+58 are similar to those found in the CEMP-s
stars that have the largest enhancements of these elements.

The excesses of C and Ba that we find for SDSS J1707+58 are much larger
than those found by Preston et al. (2006) for the RR Lyrae star TY Gru.
These excesses presumably result from the mass transfer from an AGB
companion. The referee (G.W. Preston) has pointed out that there is a
problem in reconciling how this transfer occurs with the large
separation required, because the putative AGB star must have successfully
survived RGB evolution without experiencing the consequences of Roche
Lobe overflow.

   \begin{deluxetable}{llllrccl}
   \tabletypesize{\scriptsize}
   \tablewidth{0pt}
 \tablecaption{ The Chemical Composition of SDSS J1707+58 }
   \tablehead{  
   \colhead {Element (Species)} &
   \colhead {$\log \epsilon$} &
   \colhead { [X/Fe]  } &
   \colhead { N  } &
   \colhead {$\sigma$ } 
 }
  \startdata
  C (CH)               & 8.30  & 2.79  &  1 & 0.41  \\
    Na (\ion{Na}{1})   & 5.72  & 2.40  &  2 & 0.28  \\
    Mg (\ion{Mg}{1})   & 5.80  & 1.12  &  2 & 0.27  \\
    Ca (\ion{Ca}{1})   & 4.36  & 0.94  &  2 & 0.14  \\
    Ti (\ion{Ti}{2})   & 2.70  & 0.67  &  3 & 0.19  \\
    Fe (\ion{Fe}{1})   & 4.58  & $-$2.92\tablenotemark{a} & 5  & 0.18  \\ 
    Fe (\ion{Fe}{2})   & 4.58  & $-$2.92\tablenotemark{a} & 2  & 0.20  \\ 
    Sr (\ion{Sr}{2})   & 0.70  & 0.75  &  2 & 0.65  \\
    Ba (\ion{Ba}{2})   & 2.09  & 2.83  &  5 & 0.51  \\
   \enddata
   \tablenotetext{a}{[Fe/H] values.} 
   \end{deluxetable}

\section{The Oosterhoff Types of the Carbon-Enhanced RR Lyrae Stars}

The Oosterhoff types of RR Lyraes can be found from their location in
the period $vs.$ amplitude plot (Fig 2a), and the period $vs.$ [Fe/H]
plot (Fig 2b). In the former plot, the loci of the Oo I and Oo II stars
are shown by solid and dotted lines, respectively (taken from Cacciari
et al. 2005). The parallelogram is taken from Fig. 9 of Layden et al.
(1999), and is based on the model predictions of RR Lyraes with enhanced
helium of Sweigart \& Catelan (1998); the RR Lyraes in the metal-rich
bulge globular clusters NGC 6388 and NGC 6441 lie within this
parallelogram (Pritzl et al. 2000). SDSS J1707+58 (2) lies closer to the
Oo II than the Oo I loci, and so is tentatively assigned type Oo II. TY
Gru (1) has a variable amplitude (Blazhko effect), so its Oo type is
ambiguous on this plot. Both KP Cyg (3) and UY CrB (4) lie outside the
loci of other RR Lyrae in globular clusters, and are either not RR Lyrae
stars (Andrievsky et al. 2010) or, speculatively, stars with greatly
enhanced helium. In Fig 2b, the encircled crosses show the mean periods
and [Fe/H] for the RR Lyrae stars in globular clusters: blue symbols for
Oo I, red symbols for Oo II, and green symbols for the clusters (A) NGC
104, (B) NGC 6388, and (C) NGC 6441. SDSS J1707+58 is clearly type Oo
II, TY Gru is probably type Oo II, and KP Cyg (3) and UY CrB (4) again
have a unique location, as in Fig 2a. From this evidence, we conclude
that SDSS J1707+58 is certainly of type Oo II, and TY Gru is probably of
type Oo II. The classifications of KP Cyg and UY CrB are ambiguous, and
in investigating their kinematics below, we consider absolute magnitudes
appropriate for both RR Lyrae stars (M$_{v}$ = +0.5) and BL Her stars
(M$_{v}$ = --0.25).

\section{The Kinematics of Carbon-Enhanced RR Lyrae Stars}

\begin{deluxetable}{ccccccccccccc}
\tablewidth{0cm}
\tabletypesize{\footnotesize}
\setcounter{table}{1}
\tablecaption{Data for carbon-enhanced RR Lyrae stars.}
\tablehead{ 
\colhead{ Star } &
\colhead{ Type \tablenotemark{a}     } &
\colhead{ Period      } &
\colhead{ Z \tablenotemark{b}     } &
\colhead{ [Fe/H]         } &
\colhead{ [C/Fe]                         } &
\colhead{ [Ba/Fe]                             } &
\colhead{ U \tablenotemark{c}     } &
\colhead{ V \tablenotemark{c}     } &
\colhead{ W \tablenotemark{c}     } &
\colhead{ Notes    } \\
     &      & days & kpc &   &   &   &km s$^{-1}$&km s$^{-1}$&km s$^{-1}$&  \\
}

\startdata
  SDSS 1707+58&RRab &0.678&7.59&--2.92&+2.79 &+2.83 &--224$\pm$191 &--112$\pm$114  &+195$\pm$154    & 1,A\\
  TY Gru      &RRab &0.570&4.42&--2.00&+0.89&+1.23&--075$\pm$028 &--111$\pm$032 &--043$\pm$019  &2,B \\
  KP Cyg      &RRab &0.856&0.13&+0.18 &+0.52&...  &+016$\pm$017  &+013$\pm$006  &+010$\pm$023   & 3,B \\
  KP Cyg      &CWB  &0.856&0.18&+0.18 &+0.52&     &+021$\pm$024  &+012$\pm$008  &+014$\pm$033   & 3,B  \\
  UY CrB      &RRab &0.929&1.96&--0.40&+0.65&...  &+032$\pm$021 &--051$\pm$020  &--041$\pm$015&3,B \\
  UY CrB      &CWB  &0.929&2.76&--0.40&+0.65&     &+053$\pm$031 &--063$\pm$029  &--046$\pm$021& 3,B \\
\enddata

\vspace{2mm}

\tablenotetext{a}{CWB = BL Her type Cepheid.}             
\tablenotetext{b}{Height above Galactic plane in kpc.} 
\tablenotetext{c}{Heliocentric space velocities $U$, $V$, and $W$.} 

Notes to Table 2: \\
{(1)}~~~Data from this paper.
{(2)}~~~Type, Period, and abundances from Preston et al. (2006).
{(3)}~~~Type and abundances from Wallerstein et al. (2009).     
{(A)}~~~Proper motions from the SDSS DR8 Catalog (Adelman-McCarthy et. al. 2011).
{(B)}~~~Proper motions from the UCAC3 Catalogue (Zacharias et al. 2010).
     
\end{deluxetable}

\begin{deluxetable}{ccccccccccc}
\tablewidth{0cm}
\tabletypesize{\footnotesize}
\setcounter{table}{2}
\tablecaption{Candidates for Carbon-Enhanced RR Lyrae Stars}
\tablehead{ 
\colhead{ Star } &
\colhead{ R.A. \tablenotemark{a}     } &
\colhead{ Dec. \tablenotemark{a}     } &
\colhead{ $V$ \tablenotemark{b}     } &
\colhead{ Period      } &
\colhead{ $V$-amp. \tablenotemark{c}     } &
\colhead{ $\Delta$ $\log$P  \tablenotemark{d}     } &
\colhead{ $\mid$Z$\mid$ \tablenotemark{e}     } &
\colhead{ R$_{gal}$ \tablenotemark{f}     } &
\colhead{ [C/Fe] \tablenotemark{g}       } &
\colhead{ Notes    } \\
     &      &      & (mag) &(days) &(mag) &     & (kpc)&(kpc) &       &        \\
}

\startdata
 VX Scl &023.$^{\circ}$8486&--35.$^{\circ}$1285&12.00&0.6373&0.83&+0.0344 &    2.0  & 08.4  & +1.8  & 1       \\
 BE Eri &069.$^{\circ}$5145&--01.$^{\circ}$9958&13.04&0.5796&1.00&+0.0191 &    1.6  & 10.8  & +2.4  & 2       \\
  U Lep &074.$^{\circ}$0749&--21.$^{\circ}$2172&10.58&0.5815&1.10&+0.0498 &    0.6  & 08.7  & +0.2  & 1       \\
 IV Leo &164.$^{\circ}$5523&--00.$^{\circ}$0945&15.72&0.6358&0.42&+0.0108 &    8.5   & 14.7  & +0.2 & 3        \\
 LM Leo &172.$^{\circ}$9572&--02.$^{\circ}$2405&15.51&0.6762&1.04&+0.0930 &    8.3   & 13.1  & +3.5 & 3       \\
 LO Leo &173.$^{\circ}$8448&--00.$^{\circ}$8952&15.47&0.6070&1.12&+0.0606 &    8.0  & 12.7  &  +1.1 & 3        \\
 LP Leo &174.$^{\circ}$1658&--01.$^{\circ}$4213&16.77&0.6492&0.56&--0.0122 &   15.1  & 20.0  & --0.4& 3         \\
 v370 Vir&181.$^{\circ}$5172&--02.$^{\circ}$2159&15.23&0.6993&0.89&+0.0833 &   7.5  & 11.3  & +1.7  & 3          \\
 v408 Vir&190.$^{\circ}$0147&--00.$^{\circ}$0693&16.99&0.5859&0.70&--0.0177 &  17.9   & 20.0  &+0.7 & 3          \\
 1245-04&191.$^{\circ}$2627&--04.$^{\circ}$3197&16.10&0.7460&0.36& +0.0535 &   11.2  & 13.4  & +2.4 & 4           \\
 ZZ Vir&200.$^{\circ}$9110&--04.$^{\circ}$3617&14.27&0.6841&0.89&+0.0736 &    4.6  & 07.6  & +0.9   & 1         \\
 WY Vir &203.$^{\circ}$8173&--06.$^{\circ}$9730&13.44&0.6094&1.11&+0.0603 &    3.4  & 07.0  & +3.9  & 1         \\
 1453-11&223.$^{\circ}$4947&--11.$^{\circ}$8973&15.61&0.7496&0.42&+0.0605 &    6.3  & 06.1  & +4.5  & 4         \\
 1529-05&232.$^{\circ}$4714&--05.$^{\circ}$8573&15.74&0.7343&0.46&+0.0554 &    6.5  & 05.9  & +2.2  & 4        \\
 2212-16&333.$^{\circ}$0472&--16.$^{\circ}$5797&16.27&0.6969&0.55&+0.0420 &    11.3  &12.5   & +3.4 & 4        \\
 2343+01&355.$^{\circ}$8220& +01.$^{\circ}$1742&13.04&0.6007&0.60&+0.0184 &    16.6  & 21.2  & --0.4& 4     \\
\enddata

\vspace{2mm}

\tablenotetext{a}{Epoch J2000.}             
\tablenotetext{b}{Apparent visual magnitude.  } 
\tablenotetext{c}{Amplitude of $V$ magnitude. } 
\tablenotetext{d}{Shift in $\log$ Period from locus of Oo I stars in $\log$ Period $vs$ 
 Amplitude plot. }
\tablenotetext{e}{Distance from Galactic plane in kpc.} 
\tablenotetext{f}{Galactocentric distance in kpc.} 
\tablenotetext{g}{Abundances from Christlieb et al. (2008).} 

Notes to Table 3: \\
{(1)}~~~Photometry from Kinemuchi et al. (2006).
{(2)}~~~Photometry from Schmidt \& Seth (1996).
{(3)}~~~Photometry from Vivas et al. (2004).
{(4)}~~~Photometry from Miceli  et al. (2008).
     
\end{deluxetable}

The M$_{v}$ of SDSS J1707+58 and TY Gru were taken to be +0.24 and
+0.45, respectively. We assumed [Fe/H] = --2.92 for SDSS J1707+58 and
[Fe/H] = --1.91 for TY Gru (For et al. 2011), and derived the M$_{v}$
from the following expression, given by Clementini et al. (2003): 
\begin{eqnarray}  
     M_{v} =  0.214 [Fe/H] + 0.86                 \nonumber 
\end{eqnarray}  
The interstellar extinction E$(B-V)$ for SDSS J1707+58, TY Gru, and UY
Cyg was taken from Schlafly \& Finkbeiner (2011), but not for KP Cyg,
which has too low a Galactic latitude for this method to be valid. KP
Cyg is 66 arcsec from the 7th magnitude star HD 190916, for which
Appenzeller (1996) gave E$(B-V)$ = 0.45, and a distance of 3.5 kpc. We
assumed an E$(B-V)$ of 0.32 for KP Cyg, and derived an
intensity-weighted mean $V$ magnitude for this star of 12.81 from the
light-curve of Schmidt (2002); this yields distances of 1.83 kpc and
2.59 kpc for M$_{v}$ = +0.50 and --0.25, respectively. Mean
intensity-weighted $V$ magnitudes of 15.96, 14.12, and 12.76 were
derived from the light curves of SDSS J1707+58, TY Gru, and UY Cyg (this
paper; Preston et al. 2006; Schmidt 2002, respectively). These yield
distances of 13.39 kpc and 5.32 kpc for SDSS J1707+58 and TY Gru,
respectively. The distance to UY Cyg is 2.70 kpc if it is an RR Lyrae
star, and 3.81 kpc if it is a BL Her star. Mean radial velocities of TY
Gru and KP Cyg were derived from the velocities given in Preston et al.
(2006) and Andrievsky et al. (2010), respectively. The accuracies of the
$\gamma$-velocities of KP Cyg, and to a lesser extent UY CrB, are
limited by uncertainties in their ephemerides; nevertheless, their
errors should not exceed $\pm$5 km~s$^{-1}$. A mean velocity of --39.7
km s$^{-1}$ was determined for UY CrB from three spectra obtained with
the FLWO 1.5-m telescope, using the FAST spectrograph with a 600 l/mm
grating and a 2 arcsec slit. This yielded 2.0~{\AA} resolution, with a
spectral coverage from 5500 -- 7550 \AA. Our exposures resulted in a S/N of 100
per resolution element in the continuum. The heliocentric space
velocities $U$, $V$, and $W$ were derived from the proper motions,
radial velocities, and distances using the program of Johnson \&
Soderblom (1987). These results are summarized in Table 2. TY Gru and
SDSS J1707+58 are halo stars that very probably have prograde rotations,
and so are likely to belong to the inner-halo population. KP Cyg and UY
CrB are clearly disk objects.

\begin{figure}
\includegraphics[width = 18.0 cm]{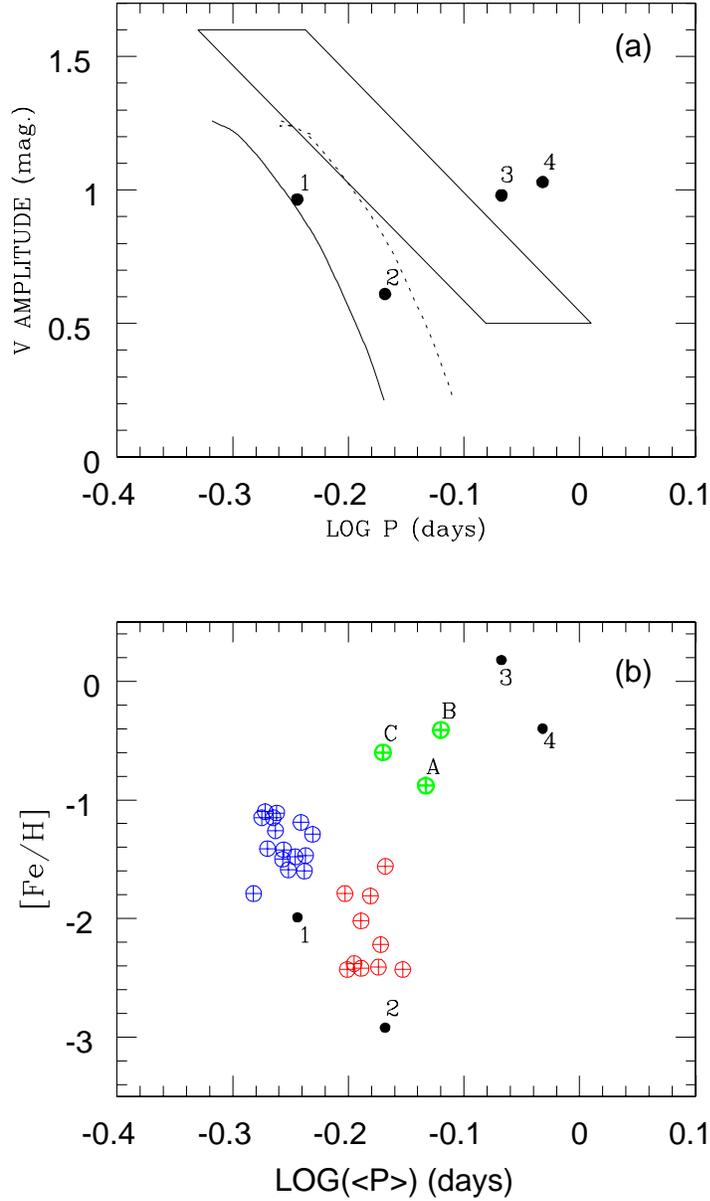}
\caption{(a) The ordinate is $V$ amplitude; the abscissa is log period (days).
 (b) The ordinate is metallicity
[Fe/H]; the abscissa is $\log$ mean period (days) for RR Lyraes in globular clusters.
Periods are from Bono et al. (2007) and [Fe/H] are from Kraft \& Ivans (2000).
 For further details, see text. }
\label{Fig2}
\end{figure}

\section{Future work}

Following Carollo et al. (2012), new CEMP RR Lyrae stars are likely to
be found among stars of low metallicity that are far from the Galactic
plane. Christlieb et al. (2008) give [C/Fe] approximate abundance ratios
for metal-poor stars found in the Hamburg/ESO objective prism survey. We
have searched this survey for RR Lyrae stars with periods greater than
0.57 days in (a) the QUEST RR Lyrae Survey (Vivas et al. 2004), (b) the
LONEOS RR Lyrae Survey (Miceli et al. 2008), and (c) the survey of RR
Lyrae stars by Wils \& Christopher (2006). We also included local RR
Lyrae stars with [Fe/H] $\leq$ --2.0 that are listed by Layden (1994).
Sixteen RR Lyrae stars were found in the Christlieb et al. (2008)
catalog; they are listed in Table 3. Twelve of these have [C/Fe] $\geq$
+0.7, and ten have [C/Fe] $\geq$ +1.0 . We call these stars
``candidates," because their abundance ratios were derived from
low-resolution objective-prism spectra. For each candidate, Table 3
gives the shift $\Delta$ $\log$ period between the $\log$ period of the
star and that of an Oo I star of the same $V$ amplitude. The six
candidates with [C/Fe] $<$ +1.0 have a mean shift of +0.016$\pm$0.018,
which shows that they are Oo I stars, whereas the ten candidates with
[C/Fe] $>$ +1.0 have a mean shift of +0.056$\pm$0.007. Thus the
candidates with [C/Fe] $>$ +1.0 primarily belong to the Oo II class
(which tend to be more metal-poor than those of the Oo I class). It is
desirable that the abundances in these stars should be determined from
high-resolution spectra, and that a more extensive search for
carbon-enhanced RR Lyrae stars should be made along the lines outlined
above (new searches of the HES prism plates for additional CEMP stars,
including possible RR Lyraes, are already underway; Placco et al. 2011).

\section{Summary}

We show that the CEMP-s star SDSS J1707+58 (studied by Aoki et al. 2008)
is the same as the RR Lyrae star VIII-14 (Suntzeff et al. 1991). We give
revised abundances for the star of [Fe/H] = --2.92, [C/Fe] = +2.79, and
[Ba/Fe] = +2.83; it is thus one of the most metal-poor RR Lyrae known,
and together with TY Gru (Preston et al. 2006), one of only two RR Lyrae
stars known to have CEMP-s properties. Both SDSS J1707+58 and TY Gru are
Oo II halo stars with prograde orbits. They differ from the mildly
carbon-enhanced stars KP Cyg and UY CrB, which are disk stars, and whose
variable type is ambiguous. Wallerstein \& Huang (2010) derived [C/Fe]
for 24 RR Lyrae stars with --2.68 $\leq$ [Fe/H] $\leq$ +0.24, and
concluded that high [C/Fe] went with metal-rich stars. We, following
Carollo et al. (2012) and others cited in the text, find that high
[C/Fe] ratios are also often associated with metal-poor RR Lyrae stars.
In support of this, we provide a list of 16 RR Lyrae stars that are
found in the catalogue of metal-poor stars of Christlieb et al. (2008);
ten of these stars have [C/Fe] $>$ +1.0, and have the characteristics of
Oo type II variables. It is important that these stars be observed at
higher spectral resolution in order to confirm their abundances.

\acknowledgments   

We thank the referee, George Preston, for his quick and thoughtful
review.

Abundance measurements are based on data collected at the Subaru
Telescope, which is operated by the National Astronomical Observatory of
Japan. We used the VizieR catalogue access tool, CDC, Strasbourg,
France. We acknowledge using data from the SDSS-III, funded by the
Alfred P. Sloan Foundation, the Participating Institutions, the NSF and
the US D.O.E.. The SDSS-III web site is http://www.sdss3.org/. Use was
also made of MAST (Multimission Archive at the STSci which is operated
for NASA by AURA), the SIMBAD database (operated at the CDS, Strasbourg,
France), ADS (the NASA Astrophysics Data System), and the arXiv e-print
server.

T.C.B. acknowledges partial funding of this work from grants PHY
02-16783 and PHY 08-22648: Physics Frontier Center/Joint Institute for
Nuclear Astrophysics (JINA), awarded by the U.S. National Science
Foundation.

\end{document}